\documentstyle{mn} 
 
\edef\resetatcatcode{\catcode`\noexpand\@\the\catcode`\@\relax}
\ifx\miniltx\undefined\else \fi
\let\miniltx\box

\def\makeatletter{\catcode`\@11\relax}
\def\makeatother{\catcode`\@12\relax}
\makeatletter

\def\@makeother#1{\catcode`#1=12\relax}

\def\@ifnextchar#1#2#3{%
  \let\reserved@d=#1%
  \def\reserved@a{#2}\def\reserved@b{#3}%
  \futurelet\@let@token\@ifnch}
\def\@ifnch{%
  \ifx\@let@token\@sptoken
    \let\reserved@c\@xifnch
  \else
    \ifx\@let@token\reserved@d
      \let\reserved@c\reserved@a
    \else
      \let\reserved@c\reserved@b
    \fi
  \fi
  \reserved@c}
\begingroup
\def\:{\global\let\@sptoken= } \:  
\def\:{\@xifnch} \expandafter\gdef\: {\futurelet\@let@token\@ifnch}
\endgroup

\def\@ifstar#1{\@ifnextchar *{\@firstoftwo{#1}}}
\long\def\@dblarg#1{\@ifnextchar[{#1}{\@xdblarg{#1}}}
\long\def\@xdblarg#1#2{#1[{#2}]{#2}}

\long\def \@gobble #1{}
\long\def \@gobbletwo #1#2{}
\long\def \@gobblefour #1#2#3#4{}
\long\def\@firstofone#1{#1}
\long\def\@firstoftwo#1#2{#1}
\long\def\@secondoftwo#1#2{#2}

\def\NeedsTeXFormat#1{\@ifnextchar[\@needsf@rmat\relax}
\def\@needsf@rmat[#1]{}
\def\ProvidesPackage#1{\@ifnextchar[%
    {\@pr@videpackage{#1}}{\@pr@videpackage#1[]}}
\def\@pr@videpackage#1[#2]{\wlog{#1: #2}}

\let\DeclareOption\@gobbletwo

\def\RequirePackage{%
  \@fileswithoptions\@pkgextension}
\def\@fileswithoptions#1{%
  \@ifnextchar[
    {\@fileswith@ptions#1}%
    {\@fileswith@ptions#1[]}}
\def\@fileswith@ptions#1[#2]#3{%
  \@ifnextchar[
  {\@fileswith@pti@ns#1[#2]#3}%
  {\@fileswith@pti@ns#1[#2]#3[]}}

\def\@fileswith@pti@ns#1[#2]#3[#4]{%
    \def\reserved@b##1,{%
      \ifx\@nil##1\relax\else
        \ifx\relax##1\relax\else
         \noexpand\@onefilewithoptions##1[#2][#4]\noexpand\@pkgextension
        \fi
        \expandafter\reserved@b
      \fi}%
      \edef\reserved@a{\zap@space#3 \@empty}%
      \edef\reserved@a{\expandafter\reserved@b\reserved@a,\@nil,}%
  \reserved@a}

\def\zap@space#1 #2{%
  #1%
  \ifx#2\@empty\else\expandafter\zap@space\fi
  #2}

\let\@empty\empty
\def\@pkgextension{sty}

\def\@onefilewithoptions#1[#2][#3]#4{%
  \input #1.#4 }

\def\typein{%
  \let\@typein\relax
  \@testopt\@xtypein\@typein}
\def\@xtypein[#1]#2{%
  \message{#2}%
  \advance\endlinechar\@M
  \read\@inputcheck to#1%
  \advance\endlinechar-\@M
  \@typein}
\def\@namedef#1{\expandafter\def\csname #1\endcsname}
\def\@nameuse#1{\csname #1\endcsname}
\def\@cons#1#2{\begingroup\let\@elt\relax\xdef#1{#1\@elt #2}\endgroup}
\def\@car#1#2\@nil{#1}
\def\@cdr#1#2\@nil{#2}
\def\@carcube#1#2#3#4\@nil{#1#2#3}
\def\@preamblecmds{}

\def\@star@or@long#1{%
  \@ifstar
   {\let\l@ngrel@x\relax#1}%
   {\let\l@ngrel@x\long#1}}

\let\l@ngrel@x\relax
\def\newcommand{\@star@or@long\new@command}
\def\new@command#1{%
  \@testopt{\@newcommand#1}0}
\def\@newcommand#1[#2]{%
  \@ifnextchar [{\@xargdef#1[#2]}%
                {\@argdef#1[#2]}}
\long\def\@argdef#1[#2]#3{%
   \@ifdefinable #1{\@yargdef#1\@ne{#2}{#3}}}
\long\def\@xargdef#1[#2][#3]#4{%
  \@ifdefinable#1{%
     \expandafter\def\expandafter#1\expandafter{%
          \expandafter
          \@protected@testopt
          \expandafter
          #1%
          \csname\string#1\expandafter\endcsname
          {#3}}%
       \expandafter\@yargdef
          \csname\string#1\endcsname
           \tw@
           {#2}%
           {#4}}}
\def\@testopt#1#2{%
  \@ifnextchar[{#1}{#1[#2]}}
\def\@protected@testopt#1{
  \ifx\protect\@typeset@protect
    \expandafter\@testopt
  \else
    \@x@protect#1%
  \fi}
\long\def\@yargdef#1#2#3{%
  \@tempcnta#3\relax
  \advance \@tempcnta \@ne
  \let\@hash@\relax
  \edef\reserved@a{\ifx#2\tw@ [\@hash@1]\fi}%
  \@tempcntb #2%
  \@whilenum\@tempcntb <\@tempcnta
     \do{%
         \edef\reserved@a{\reserved@a\@hash@\the\@tempcntb}%
         \advance\@tempcntb \@ne}%
  \let\@hash@##%
  \l@ngrel@x\expandafter\def\expandafter#1\reserved@a}
\long\def\@reargdef#1[#2]#3{%
  \@yargdef#1\@ne{#2}{#3}}
\def\renewcommand{\@star@or@long\renew@command}
\def\renew@command#1{%
  {\escapechar\m@ne\xdef\@gtempa{{\string#1}}}%
  \expandafter\@ifundefined\@gtempa
     {\@latex@error{\string#1 undefined}\@ehc}%
     {}%
  \let\@ifdefinable\@rc@ifdefinable
  \new@command#1}
\long\def\@ifdefinable #1#2{%
      \edef\reserved@a{\expandafter\@gobble\string #1}%
     \@ifundefined\reserved@a
         {\edef\reserved@b{\expandafter\@carcube \reserved@a xxx\@nil}%
          \ifx \reserved@b\@qend \@notdefinable\else
            \ifx \reserved@a\@qrelax \@notdefinable\else
              #2%
            \fi
          \fi}%
         \@notdefinable}
\let\@@ifdefinable\@ifdefinable
\long\def\@rc@ifdefinable#1#2{%
  \let\@ifdefinable\@@ifdefinable
  #2}
\def\newenvironment{\@star@or@long\new@environment}
\def\new@environment#1{%
  \@testopt{\@newenva#1}0}
\def\@newenva#1[#2]{%
   \@ifnextchar [{\@newenvb#1[#2]}{\@newenv{#1}{[#2]}}}
\def\@newenvb#1[#2][#3]{\@newenv{#1}{[#2][#3]}}
\def\renewenvironment{\@star@or@long\renew@environment}
\def\renew@environment#1{%
  \@ifundefined{#1}%
     {\@latex@error{Environment #1 undefined}\@ehc
     }{}%
  \expandafter\let\csname#1\endcsname\relax
  \expandafter\let\csname end#1\endcsname\relax
  \new@environment{#1}}
\long\def\@newenv#1#2#3#4{%
  \@ifundefined{#1}%
    {\expandafter\let\csname#1\expandafter\endcsname
                         \csname end#1\endcsname}%
    \relax
  \expandafter\new@command
     \csname #1\endcsname#2{#3}%
     \l@ngrel@x\expandafter\def\csname end#1\endcsname{#4}}

\def\providecommand{\@star@or@long\provide@command}
\def\provide@command#1{%
  {\escapechar\m@ne\xdef\@gtempa{{\string#1}}}%
  \expandafter\@ifundefined\@gtempa
    {\def\reserved@a{\new@command#1}}%
    {\def\reserved@a{\renew@command\reserved@a}}%
   \reserved@a}%

\def\@ifundefined#1{%
  \expandafter\ifx\csname#1\endcsname\relax
    \expandafter\@firstoftwo
  \else
    \expandafter\@secondoftwo
  \fi}

\chardef\@xxxii=32
\mathchardef\@Mi=10001
\mathchardef\@Mii=10002
\mathchardef\@Miii=10003
\mathchardef\@Miv=10004

\newcount\@tempcnta
\newcount\@tempcntb
\newif\if@tempswa\@tempswatrue
\newdimen\@tempdima
\newdimen\@tempdimb
\newdimen\@tempdimc
\newbox\@tempboxa
\newskip\@tempskipa
\newskip\@tempskipb
\newtoks\@temptokena

\long\def\@whilenum#1\do #2{\ifnum #1\relax #2\relax\@iwhilenum{#1\relax
     #2\relax}\fi}
\long\def\@iwhilenum#1{\ifnum #1\expandafter\@iwhilenum
         \else\expandafter\@gobble\fi{#1}}
\long\def\@whiledim#1\do #2{\ifdim #1\relax#2\@iwhiledim{#1\relax#2}\fi}
\long\def\@iwhiledim#1{\ifdim #1\expandafter\@iwhiledim
        \else\expandafter\@gobble\fi{#1}}
\long\def\@whilesw#1\fi#2{#1#2\@iwhilesw{#1#2}\fi\fi}
\long\def\@iwhilesw#1\fi{#1\expandafter\@iwhilesw
         \else\@gobbletwo\fi{#1}\fi}
\def\@nnil{\@nil}
\def\@empty{}
\def\@fornoop#1\@@#2#3{}
\long\def\@for#1:=#2\do#3{%
  \expandafter\def\expandafter\@fortmp\expandafter{#2}%
  \ifx\@fortmp\@empty \else
    \expandafter\@forloop#2,\@nil,\@nil\@@#1{#3}\fi}
\long\def\@forloop#1,#2,#3\@@#4#5{\def#4{#1}\ifx #4\@nnil \else
       #5\def#4{#2}\ifx #4\@nnil \else#5\@iforloop #3\@@#4{#5}\fi\fi}
\long\def\@iforloop#1,#2\@@#3#4{\def#3{#1}\ifx #3\@nnil
       \expandafter\@fornoop \else
      #4\relax\expandafter\@iforloop\fi#2\@@#3{#4}}
\def\@tfor#1:={\@tf@r#1 }
\long\def\@tf@r#1#2\do#3{\def\@fortmp{#2}\ifx\@fortmp\space\else
    \@tforloop#2\@nil\@nil\@@#1{#3}\fi}
\long\def\@tforloop#1#2\@@#3#4{\def#3{#1}\ifx #3\@nnil
       \expandafter\@fornoop \else
      #4\relax\expandafter\@tforloop\fi#2\@@#3{#4}}
\long\def\@break@tfor#1\@@#2#3{\fi\fi}
\def\@removeelement#1#2#3{%
  \def\reserved@a##1,#1,##2\reserved@a{##1,##2\reserved@b}%
  \def\reserved@b##1,\reserved@b##2\reserved@b{%
    \ifx,##1\@empty\else##1\fi}%
  \edef#3{%
    \expandafter\reserved@b\reserved@a,#2,\reserved@b,#1,\reserved@a}}

\let\ExecuteOptions\@gobble

\def\@latex@error#1#2{%
  \errhelp{#2}\errmessage{#1}}

\bgroup\uccode`\!`\%\uppercase{\egroup
\def\@percentchar{!}}

\ifx\@@input\@undefined
 \let\@@input\input
\fi

\def\input{\@ifnextchar\bgroup\@iinput\@@input}
\def\@iinput#1{\input{#1} }

    \def\filename@parse#1{%
      \let\filename@area\@empty
      \expandafter\filename@simple#1.\\}

  \def\filename@simple#1.#2\\{%
    \ifx\\#2\\%
       \let\filename@ext\relax
    \else
       \edef\filename@ext{\filename@dot#2\\}%
    \fi
    \edef\filename@base{#1}}
  \def\filename@dot#1.\\{#1}

\long\def \IfFileExists#1#2#3{%
  \openin\@inputcheck#1 %
  \ifeof\@inputcheck
    \ifx\input@path\@undefined
      \def\reserved@a{#3}%
    \else
      \def\reserved@a{\@iffileonpath{#1}{#2}{#3}}%
    \fi
  \else
    \closein\@inputcheck
    \edef\@filef@und{#1 }%
    \def\reserved@a{#2}%
  \fi
  \reserved@a}
\long\def\@iffileonpath#1{%
  \let\reserved@a\@secondoftwo
  \expandafter\@tfor\expandafter\reserved@b\expandafter
             :\expandafter=\input@path\do{%
    \openin\@inputcheck\reserved@b#1 %
    \ifeof\@inputcheck\else
      \edef\@filef@und{\reserved@b#1 }%
      \let\reserved@a\@firstoftwo%
      \closein\@inputcheck
      \@break@tfor
    \fi}%
  \reserved@a}
\long\def \InputIfFileExists#1#2{%
  \IfFileExists{#1}%
    {#2\@addtofilelist{#1}\@@input \@filef@und}}

\chardef\@inputcheck0

\let\@addtofilelist \@gobble

\def\@defaultunits{\afterassignment\remove@to@nnil}
\def\remove@to@nnil#1\@nnil{}

\newdimen\leftmarginv
\newdimen\leftmarginvi

\newdimen\@ovxx
\newdimen\@ovyy
\newdimen\@ovdx
\newdimen\@ovdy
\newdimen\@ovro
\newdimen\@ovri
\newdimen\@xdim
\newdimen\@ydim
\newdimen\@linelen
\newdimen\@dashdim

\long\def\mbox#1{\leavevmode\hbox{#1}}

\let\@onlypreamble\@gobble

\let\protect\relax

\newdimen\fboxsep
\newdimen\fboxrule

\def\@height{height} \def\@depth{depth} \def\@width{width}
\def\@minus{minus}
\def\@plus{plus}
\def\hb@xt@{\hbox to}

\long\def\@begin@tempboxa#1#2{%
   \begingroup
     \setbox\@tempboxa#1{\color@begingroup#2\color@endgroup}%
     \def\width{\wd\@tempboxa}%
     \def\height{\ht\@tempboxa}%
     \def\depth{\dp\@tempboxa}%
     \let\totalheight\@ovri
     \totalheight\height
     \advance\totalheight\depth}
\let\@end@tempboxa\endgroup

\let\set@color\relax
\let\color@begingroup\relax
\let\color@endgroup\relax
\let\color@setgroup\relax

\let\color@hbox\relax
\let\color@vbox\relax
\let\color@endbox\relax

\begingroup
  \catcode`P=12
  \catcode`T=12
  \lowercase{
    \def\x{\def\rem@pt##1.##2PT{##1\ifnum##2>\z@.##2\fi}}}
  \expandafter\endgroup\x
\def\strip@pt{\expandafter\rem@pt\the}

\def\@input#1{%
  \IfFileExists{#1}{\@@input\@filef@und}{\message{No file #1.}}}

\def\@warning{\immediate\write16}

\def\Gin@driver{dvips.def}
\input graphicx.sty

\resetatcatcode

\newif\ifAMStwofonts 
 
\ifoldfss 
  \ifCUPmtlplainloaded \else 
    \NewTextAlphabet{textbfit} {cmbxti10} {} 
    \NewTextAlphabet{textbfss} {cmssbx10} {} 
    \NewMathAlphabet{mathbfit} {cmbxti10} {} 
    \NewMathAlphabet{mathbfss} {cmssbx10} {} 
  \fi 
  \ifAMStwofonts 
    \ifCUPmtlplainloaded \else 
      \NewSymbolFont{upmath} {eurm10} 
      \NewSymbolFont{AMSa} {msam10} 
      \NewMathSymbol{\upi}     {0}{upmath}{19} 
      \NewMathSymbol{\umu}     {0}{upmath}{16} 
      \NewMathSymbol{\upartial}{0}{upmath}{40} 
      \NewMathSymbol{\leqslant}{3}{AMSa}{36} 
      \NewMathSymbol{\geqslant}{3}{AMSa}{3E}

    \fi 
  \fi 
\fi 
 
\ifnfssone 
  \newmathalphabet{\mathit} 
  \addtoversion{normal}{\mathit}{cmr}{m}{it} 
  \addtoversion{bold}{\mathit}{cmr}{bx}{it} 
  \newmathalphabet{\mathbfit} 
  \addtoversion{normal}{\mathbfit}{cmr}{bx}{it} 
  \addtoversion{bold}{\mathbfit}{cmr}{bx}{it} 
  \newmathalphabet{\mathbfss} 
  \addtoversion{normal}{\mathbfss}{cmss}{bx}{n} 
  \addtoversion{bold}{\mathbfss}{cmss}{bx}{n} 
  \ifAMStwofonts 
    \ifCUPmtlplainloaded \else 
      \UseAMStwoboldmath 
      \makeatletter 
      \new@mathgroup\upmath@group 
      \define@mathgroup\mv@normal\upmath@group{eur}{m}{n} 
      \define@mathgroup\mv@bold\upmath@group{eur}{b}{n} 
      \edef\UPM{\hexnumber\upmath@group} 
      \new@mathgroup\amsa@group 
      \define@mathgroup\mv@normal\amsa@group{msa}{m}{n} 
      \define@mathgroup\mv@bold\amsa@group{msa}{m}{n} 
      \edef\AMSa{\hexnumber\amsa@group} 
      \makeatother 
      \mathchardef\upi="0\UPM19 
      \mathchardef\umu="0\UPM16 
      \mathchardef\upartial="0\UPM40 
      \mathchardef\leqslant="3\AMSa36 
      \mathchardef\geqslant="3\AMSa3E 
    \fi 
  \fi 
\fi 
 
\ifnfsstwo 
  \DeclareMathAlphabet{\mathbfit}{OT1}{cmr}{bx}{it} 
  \SetMathAlphabet\mathbfit{bold}{OT1}{cmr}{bx}{it} 
  \DeclareMathAlphabet{\mathbfss}{OT1}{cmss}{bx}{n} 
  \SetMathAlphabet\mathbfss{bold}{OT1}{cmss}{bx}{n} 
  \ifAMStwofonts 
    \ifCUPmtlplainloaded \else 
      \DeclareSymbolFont{UPM}{U}{eur}{m}{n} 
      \SetSymbolFont{UPM}{bold}{U}{eur}{b}{n} 
      \DeclareSymbolFont{AMSa}{U}{msa}{m}{n} 
      \DeclareMathSymbol{\upi}{0}{UPM}{"19} 
      \DeclareMathSymbol{\umu}{0}{UPM}{"16} 
      \DeclareMathSymbol{\upartial}{0}{UPM}{"40} 
      \DeclareMathSymbol{\leqslant}{3}{AMSa}{"36} 
      \DeclareMathSymbol{\geqslant}{3}{AMSa}{"3E} 
    \fi 
  \fi 
\fi 
 
\ifCUPmtlplainloaded \else 
  \ifAMStwofonts \else 
    \def\upi{\pi} 
    \def\umu{\mu} 
    \def\upartial{\partial} 
  \fi 
\fi

\def\simlt{\lower.5ex\hbox{$\; \buildrel < \over \sim \;$}}
\def\simgt{\lower.5ex\hbox{$\; \buildrel > \over \sim \;$}}

\def\mp{M$_{\odot}$ pc$^{-2}$}
\def\mg{M$_{\odot}$ pc$^{-2}$ Gyr$^{-1}$}

\def\l{$\lambda$}

\def\HI{\mbox{H\,{\sc i}}}

\begin{document}

\title{Chemical and spectrophotometric evolution of Low Surface Brightness galaxies  }

\author[S. Boissier, D. Monnier Ragaigne, N. Prantzos, W. van Driel, C. Balkowski, K. O'Neil]
       {S. Boissier$^{1,5}$, D. Monnier Ragaigne$^2$, N. Prantzos$^3$, W. van Driel$^2$, C. Balkowski$^2$, 
\newauthor
K. O'Neil$^4$ \\
1) Institute of Astronomy, University of Cambridge, Madingley Road, Cambridge. CB3 0HA, United Kingdom. \\
2) Observatoire de Paris, Section de Meudon, GEPI, CNRS FRE 2459 and Universit\'e de Paris7, 
 5 place Jules Janssen, \\ 92195 Meudon, France \\ 
3) Institut d'Astrophysique de Paris, 98bis Bd. Arago, 75014 Paris, France \\
4) Arecibo observatory, HC03, Box 53995, Arecibo, PR 00612, U.S.A. \\
5) Carnegie Observatories, 813 Santa Barbara Street, Pasadena, California 91101, U.S.A. \\
}

\date{ }

\pagerange{\pageref{firstpage}--\pageref{lastpage}}
\pubyear{2002}
\maketitle

\label{firstpage}

\begin{abstract}
 Based on the results of recent surveys, we have constructed a
relatively homogeneous set of observational data concerning the
chemical and photometric properties of Low Surface Brightness galaxies
(LSBs).  
We have compared the properties of this data set with the
predictions of models of the chemical and spectrophotometric evolution
of LSBs.  The basic idea behind the models, i.e. that LSBs are similar
to 'classical' High Surface Brightness spirals except for a larger
angular momentum, is found to be consistent with the results of their
comparison with these data.  However, some observed properties of the
LSBs (e.g. their colours, and specifically the existence of red LSBs)
as well as the large scatter in these properties, cannot be reproduced
by the simplest models with smoothly evolving star formation rates
over time. We argue that the addition of bursts and/or truncations in
the star formation rate histories can alleviate that discrepancy.
\end{abstract}

\begin{keywords}
Galaxies: general - evolution - spiral - photometry - stellar content  
\end{keywords}

\section{Introduction}

Galaxies with a central  disc surface brightness (in the $B$ band)
well below the Freeman value of  $\mu_{B,0}$=21.65 mag arcsec$^{-2})$ 
- typical of the average previously catalogued ``classical'' High Surface 
Brightness  galaxies or HSBs - are classified as LSBs. 
In the last decade, in particular,  a considerable body of observational data
has been collected on those objects
(e.g. de Blok, van der Hulst \& Bothun 1995 ; 
O'Neil, Bothun \& Cornell 1997a ; 
O'Neil et al. 1997b ;
Beijersbergen, de Blok \& van der Hulst 1999 ;
O'Neil, Bothun \& Schombert 2000a ; 
van den Hoek et al. 2000 ;
Matthews, van Driel \& Monnier Ragaigne 2001; Galaz et al. 2002).
Although there is no unambiguous definition of an LSB galaxy,
in the following we will adopt as the limit between HSBs and LSBs a
central disc surface brightness value of $\mu_{B,0}$=22 mag arcsec$^{-2}$, 
a commonly used criterion.

LSB galaxies may turn out to be crucial for 
studies of galaxy formation and evolution and of
the `cosmic' chemical evolution  if they constitute the 
majority of galaxies (as suggested by O'Neil \& Bothun 2000).
They are also of interest in the field of high redshift quasar
absorbers: some Damped Lyman Alpha systems (DLAs) are  identified with
LSBs, and DLAs could be due as much to LSBs as to `classical' spirals
(e.g. Boissier, P\'eroux \& Pettini 2003 and references therein).

Observations have shown a large variety in the properties of
of LSB galaxies, which range from dwarfs to massive systems
(e.g. Bothun, Impey \& McGaugh 1997) and 
from the ``blue'' to the ``red'' edge of galactic colours 
(O'Neil et al. 2000a). This variety suggests that they form a heterogeneous
family, as argued by Bell et al. (2000).

Nevertheless, their potentially interesting implications for galactic
astrophysics motivated various theoretical studies, aiming towards an
understanding of their nature. For instance, Gerritsen \& de Blok
(1999) used N-body simulations to study their star formation, while
van den Hoek et al. (2000) compiled observational characteristics of
24 LSBs and tried to derive their star formation histories by
comparison with simplified chemical evolution models. On the other
hand, Jimenez et al. (1998) modelled the chemical and
spectro-photometric evolution of LSBs by assuming that they are discs
with larger angular momentum than HSBs; this idea had been suggested
through the study of the formation of disc galaxies by Dalcanton,
Spergel \& Summers (1997), who considered a gravitationally
self-consistent model for the formation of both LSB and HSB discs.

In a recent series of papers (Boissier \& Prantzos 2000, Prantzos and
Boisier 2000, Boissier and Prantzos 2001, Boissier et al. 2001) the
chemical and spectrophotometric evolution of spiral galaxies of
various masses and spin parameters was computed and compared quite
successfully to a large number of observational data: Tully-Fisher
relation, colours, integrated spectra, star formation efficiency, gas
fraction etc.  Those works showed that for relatively large values of
the spin parameter $\lambda$ the properties of discs ressembled
closely those of LSBs (in particular the central surface brightness).

Following the ideas of Dalcanton et al. (1997) and Jimenez et
al. (1998) in the present paper we extend the same models to larger
values of the spin parameter, keeping the remaining galactic physics
(stellar Initial Mass Function, prescriptions for Star Formation Rate
and Infall rate etc.) the same.  We compare then our results to an
extended set of observed properties of LSBs (\HI\ mass, colours,
abundances).  Our main motivation is to answer the question: can HSB
spirals and LSBs be modelled in the same framework, where LSBs merely
have a larger angular momentum, or is it necessary to invoke specific
assumptions in order to explain their properties?

In section \ref{secobs} we present the compilation of data on LSBs
properties that we adopted for our study. In section \ref{models} we
explain how the simplest (1st order) models of LSBs are constructed on
the basis of our - quite successful - HSBs models, by merely
increasing the spin parameter. These simple models are then compared
to the observations in section \ref{seccomp}.  In section
\ref{burstanti} bursts and truncations are added to the smooth star
formation history of the ``simple models'', in order to obtain better
fits to the ensemble of the observational data.  Section
\ref{findesharicots} gives a summary of the obtained results.

\section{Low Surface Brightness galaxies: the  data sample}

\label{secobs}

Low Surface Brightness galaxies (LSBs) display a very broad range of
sizes, masses and spectral line widths. Their typical observed 21 cm
\HI\ line width is about 100 km/s, though it can be as high as 600
km/s. ``Giant'' LSBs have also very large scalelengths, the most
striking case being Malin-1.  In some sense, their large velocities
and sizes make them similar to HSB spirals, as argued in Bothun et
al. (1997).

In this paper, we explore possible links in the physical properties/histories
between HSB and LSB discs. Since we have no 
{\it a priori} idea of which of the LSBs could actually be linked to HSBs, we 
consider observations for the whole range of LSBs, from dwarfs to giants.
Our sample consists of the following data sets, available in the literature.

\underline{O'Neil et al. LSBs sample}

The observational data on LSBs from O'Neil et al. (1997a,b) provides
information on the surface brightness, scalelength, absolute magnitude of the disc
and colours for 127 galaxies.
80\% of them are well fitted by an exponential profile, which justifies the use of a 
disc model for studying LSBs. The data suggest that a large fraction of LSBs 
are redder than previously thought. \HI\ data for some of these galaxies
are given in O'Neil, Bothun, \& Schombert (2000) and Chung et al. (2002).
Note that some of these objects are members of galaxy 
clusters (Pegasus and Cancer). 

\underline{de Blok et al. LSBs sample}

For 21 late-type field LSBs from the sample of Schombert \& Bothun (1988) 
and the UGC, de Blok et al. (1995) published disc scalelengths and absolute magnitudes 
and colours; for subsets of that sample \HI\ data are given in de Blok et al. (1996), 
abundances in de Blok \& van der Hulst (1998a), CO line upper limits
in de Blok \& van der Hulst (1998b) and H$\alpha$ rotation curves in 
McGaugh, Rubin \& de Blok (2001).
The gas content of those galaxies is also available from the work of
van den Hoek et al. (2000).

\underline{Bulge-dominated LSBs sample}

Although most of the LSBs seem to be late-type objects without any
apparent bulge, some of them clearly do have bulges, and we do not 
exclude these from our analysis, to avoid introducing a bias against the most
evolved LSB discs. For this reason, we also use the data on 
surface brightness, scalelength, and colours of 20 galaxies from 
Beijersbergen et al. (1999). 
As our models concern only the disc component of LSBs, we used the ``area-weighted''
colours given by Beijersbergen et al., which should be representative of the discs.

\underline{Giant LSBs sample}

We also use the data of Matthews et al. (2001), which provides surface photometry 
and \HI\ data for 16 giant LSB galaxies, with high luminosities 
($\sim$10$^{10}$ L$_{\odot}$)
and large scalelengths ($>6$ kpc). Though LSB Giants are relatively rare, they
may play an important role in our investigation of the link between HSBs and
LSBs because of their similar sizes/velocities (note that there are many more HSBs with 
large velocity widths than with small ones, like for LSBs).

\underline{Infra-red selected LSBs sample}

A new sample of LSBs selected from  the 2MASS near-infrared survey
(Monnier Ragaigne et al., 2002, MR2002) is used as well. 
Though these 4,000 objects were selected on their low $K_s$-band central
disc surface brightness ($\mu_{K,0}$$<$18 mag arcsec$^{-2}$) they turned out to
be bluer than expected; the $B$-band central disc surface brightnesses
of the 20 objects for which we have optical surface photometry actually lie in 
the range of $\sim$21-22.5  mag arcsec$^{-2}$, straddling the 
22  mag arcsec$^{-2}$ $B$-band selection criterion for LSBs we adopted in the 
present paper.
Here we use the subset of 20 galaxies for which we presently have available
$BVRI$ surface photometry, \HI\ spectra and $JHK$ photometry.
This sample allows the exploration of the link between optically HSB and
LSB galaxies.

\underline{Homogenisation of the data}

Because the data sets have different origins, they need homogenisation before
a proper comparison with our disc models. We adopted the following procedure:

$\bullet$ All distance-dependent quantities are reduced to distances based on 
a Hubble constant of 65 km/s/Mpc. 

$\bullet$ The models presented in the next section were  computed
under the hypothesis that the discs are observed face-on, and the data were 
corrected accordingly to an inclination angle i=0$^{\circ}$. 
This affects magnitudes and surface 
brightness by a quantity equal to $-2.5 C \log(\cos i)$, where 
$C$ ranges from 0 for an optically thick disc to 1 for an optically thin 
one. Since LSBs are usually not much evolved chemically (they have low metallicities), 
it is likely that they are dust-free and we assume $C$=1. This assumption
is justified by the fact that the colours of LSBs seem to be
independent of inclination  (e.g. O'Neil et al. 1997a).

\label{secmagalpha}
$\bullet$ The absolute total disc magnitude used in this paper, $M_{B,d}$,
is deduced from the fit with an exponential profile (e.g. O'Neil et al. 1997a), i.e.
$M_{B,d}=\mu_0-2.5 \log(2 \pi R_d^2)$, where $\mu_0$ and $R_d$ 
are the central surface brightness and the scalelength, in
mag arcsec$^{-2}$ and arcsec, respectively.

$\bullet$ The total gas mass is obtained by correcting the \HI\ mass
for the helium fraction ($M_G$/$M_{HI}$=1.4). The molecular component
in LSBs is generally negligible, though a number of LSBs have now been detected
in CO (Matthews \& Gao 2001; O'Neil, Hofner \& Schinnerer 2000, and
references therein).

\section{Models: from ``classical'' HSB spirals to LSBs}

\label{models}

Dalcanton et al. (1997) suggested that LSBs could be the large angular momentum 
equivalents of ``classical'' spirals. That idea was adopted in Jimenez et al. 
(1998) who constructed
models relatively similar to those used in the present study; they 
showed that this assumption can lead to reasonably good results, at least 
for the limited observational sample they used (which did not include 
objects like the red LSBs  of the O'Neil et al. sample).

In this paper, we have adopted the same assumption, i.e. that LSBs can
be explained with models similar to those of HSB spirals, but with
a larger angular momentum.  The strength of this work is that it relies
on detailed models of the chemical and spectro-photometric evolution
of disc galaxies, already shown to be successful in reproducing the
main properties of the Milky Way and of nearby HSB spirals: see
Boissier \& Prantzos 1999 (BP99) for the Milky Way model, Boissier \&
Prantzos 2000 (BP2000) for the extension to other spirals and Boissier
et al. (2001) on the agreement of the models with the observed gas
fraction, star formation efficiency and ages of nearby spirals.

Notice that Schombert, McGaugh \& Eder (2001) determined the gas
fraction of LSB dwarf galaxies, and argued that their data were
compatible with the BP2000 models of discs with the largest angular
momenta. As an alternative possibility, they also suggested that LSBs
could be on average younger than HSBs.  That conclusion is also in
agreement with our models, since large angular momentum galaxies
(LSBs) form the bulk of their stars later, on average, and thus they
appear younger than HSBs.  We shall note that their galaxies are
dwarf-LSBs, corresponding to the lower masses found in 
the samples we collected (section 2).

Since the LSBs models presented here are a simple extrapolation of
those presented previously in
this series, we will recall here only briefly their main features (section \ref{generalmod});
we discuss,  in particular, the role of the spin parameter and of
its distribution (section \ref{spinmod}), since  we assume that it is at the origin
of the differences between HSBs and LSBs.

\subsection{General features}
 
\label{generalmod}

In BP99, the disc of our Galaxy is simulated as an ensemble of concentric rings 
gradually built up by infall of gas of primordial composition. The adopted 
Star Formation Rate (SFR), as suggested by the density wave theory, is
\begin{equation}
\label{equasfr}
\psi = \alpha \Sigma_{gas}^{1.5} \frac{V(R)}{R}
\end{equation}
where V(R) is the disc rotation velocity as function of radius,
$\Sigma_{gas}$ the gas density (in \mp),
$\psi$ the star formation rate (in \mg), 
and $\alpha$ is an efficiency parameter adjusted as to reproduce the properties of the
Solar Neighborhood. That form of the SFR
is in agreement with the observed current SFR profile of the Milky Way (BP1999) and of 
some nearby spiral galaxies (Boissier et al. 2003, in preparation).
We adopt the Initial Mass Function (IMF) of Kroupa et al. (1993), characterised by a
flattening at low masses, and consider it as universal, i.e. not evolving in time or space.
The infall rate is declining exponentially, with a timescale $\tau$ = 7 Gyr in the local disc
in order to reproduce the observed G-dwarf metallicity distribution. 
Inner galactic zones are formed earlier than the outer disc (``inside-out'' disc formation). 
The evolution is computed independently at each radius and the global properties
(masses, luminosities, colours) are obtained by integration or by fitting the obtained profiles
to recover scalelengths and central surface brightnesses.
For further details on the  ingredients and the model of the Milky Way, see BP99.

In BP2000, the mass and scalelength of the discs of HSB spirals are  computed with
similar models, using  simple ``scaling laws'', as suggested by Mo, Mao \& White (1998) 
in the framework of  Cold Dark Matter scenarios of galaxy formation.
Within this simplified approach, discs are characterised by two parameters: their circular 
velocity $V_C$ (determined by the mass of the dark halo), and their spin parameter 
$\lambda=J |E|^{1/2}G^{-1}M^{-5/2}$
where $J$, $E$, $M$ are, respectively, the angular momentum, total energy and 
mass of the dark halo, whereas $\lambda$ is a dimensionless quantity measuring the specific 
angular momentum of the dark halo.
\label{secscal}
The disc scalelength can then be expressed as:
\begin{equation}
\label{equascal1}
R_d = R_{d,MW} \times \frac{V_C}{V_{C,MW}} \times \frac{\lambda}{\lambda_{MW}}
\end{equation}
and the central surface density as 
\begin{equation}
\label{equascal2}
\Sigma_0=\Sigma_{0,MW} \times \frac{V_C}{V_{C,MW}} \times 
\left(\frac{\lambda}{\lambda_{MW}}\right)^{-2}
\end{equation}
where the index $MW$ refers to the corresponding value in the Milky
Way (see BP2000 for details). The disc mass varies with $V_C^3$.  From
equation \ref{equascal2} it is obvious that the central surface
density depends strongly on the spin parameter and that large values
of it are likely to represent LSB galaxies.  Actually, Dekel and Woo
(2002) present some observational evidences (based on the Sloan
Digital Sky Survey data) that the surface brightness decreases with
the spin parameter $\lambda$, even if they consider $\lambda$ as a
``second parameter'' for the LSB class of galaxies, the first
parameter being the stellar mass. This is not surprising since they
include all dwarf galaxies in their study, and have then a much larger
range of stellar mass than we consider. Note that our surface brightnesses
also depends on the stellar mass (what can be deduced from figure \ref{figmur}).

On the other hand, since the same star formation rate prescription  (equation \ref{equasfr}) is
applied for all galaxies, a lower star formation rate in LSBs will result naturally 
from the dependence of the SFR on radius and on gas surface density.

In BP2000 it was argued that observational data on nearby spirals
suggest  that the infall 
time-scale decreases with galaxy mass, i.e. that less massive discs
were formed later, on average; this   dependence of infall time-scales 
on the circular velocity $V_C$ is also included here. Infall rates depend 
also on the local surface density (in order to produce an inside-out formation of 
the galactic discs) and this results in a slower formation of LSB galaxies. 

Because LSBs display a large range of \HI\ line widths, 
we computed a new set of models with six values for the 
circular velocity:  40, 80, 150, 220, 290 and 360 km/s
(only the first one was not considered in BP2000).

\subsection{Spin parameter distribution}

In BP2000, models of HSB spirals were obtained for  80 $< V_C {\rm (km/s)} <$ 360
and  $0.02 \simlt \lambda \simlt 0.07$. It was noted that for $ 0.07 \simlt \lambda \simlt 0.09$,
the surface densities were so low that the discs were in fact at the limit of 
LSB galaxies ($\mu_{B,0}$ between 22.5 and 23 mag arcsec$^{-2}$).

The models of BP2000 with $\lambda \simgt 0.06$ have a central surface
brightness close to 22 mag arcsec$^{-2}$. In this work we consider
larger values of the spin parameter $\lambda$ that produce
``naturally'' LSB galaxies.
\label{spinmod}
\begin{figure}  
\includegraphics[angle=-90,width=0.5\textwidth]{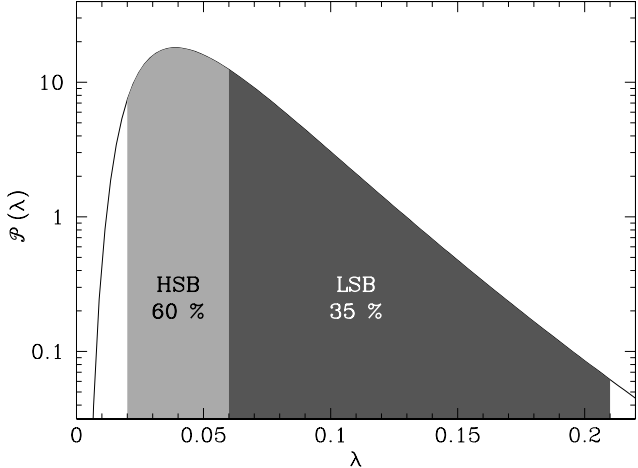}
\caption{Distribution of the spin parameter $\lambda$ 
(measuring the specific angular momentum of galaxies), as obtained in
N-body simulations (equation \ref{equalam}, section \ref{spinmod}).
If the high surface brightness spirals correspond to
$0.02<\lambda<0.06$, and the low surface brightness galaxies to
$\lambda>0.06$ (as suggested by the models of BP2000), then the former
would represent 60 \% of the total number of disc galaxies, and the
latter 35 \%.
\label{figlam}}
\end{figure}

Mo et al. (1998) provide a largely used spin parameter distribution 
(figure \ref{figlam})
which gives a reasonable fit to the results of numerical simulations
of the formation of dark matter halos; it can be expressed as follows:
\begin{equation}
\label{equalam}
P_{\lambda}(\lambda)d\lambda \ = \ \frac{1}{\sqrt{2\pi}\sigma_{\lambda}} \
exp \left[ - \frac{ln^2(\lambda/\bar{\lambda})}{2 \sigma_{\lambda}^2} 
\right]\frac{d\lambda}{\lambda}
\end{equation}
where the average spin parameter $\bar{\lambda}$ = 0.05 and $\sigma_{\lambda}$ = 0.5.

Values of \l{} larger than 0.06 will produce discs with a central blue
surface brightness lower than 22. mag arcsec$^{-2}$. According to the
distribution shown in figure \ref{figlam}, 35\% of the galaxies will
have spin parameters in the interval [0.06,0.21], while the 60\% in
the interval [0.02,0.06] are likely to be HSB spirals (see BP2000).
Note that the limits of the intervals are a bit arbitrary since the 
surface brightness depends slightly on the velocity $V_C$ too. 
Especially, galaxies with the lower rotational velocities can be LSBs
even with lower spin parameters.
The limits adopted here represent however a conservative estimate for
the number of LSB galaxies.

These figures are not fully consistent with the claim by O'Neil \&
Bothun (2000) that ``the majority of the galaxies'' are LSBs.  However
they suggest that a significant part of the population of LSB galaxies
could indeed be discs of large spins.  They are also at odds with the
claim of Dekel and Woo (2002) that 95 \% of galaxies are LSBs, but
this is mainly due to differences in definitions.  Dekel and Woo
(2002) include in their ``LSB'' class all dwarf galaxies, while we limit
ourselves to discs with rotation velocities larger than 40 km/s. This
choice excludes of course a large number of galaxies because of the
shape of the luminosity function and is justified by the fact that we
are indeed interested in the behaviours of (relatively) 
massive LSB disc galaxies, comparing them to similar HSBs.

The LSB disc models constructed for this paper were therefore computed
for 4 values of $\lambda$: 0.07, 0.09, 0.15 and 0.21 (the first two
values were already used in BP2000).  All other ingredients of the
models are identical to those used in the HSB disc models, as
discussed in section 3.1.

The present work can not be considered as a proof that LSB galaxies
are large-spin parameter discs since we did considerate other
possibilities as for instance that LSB galaxies are settling in halos
with larger baryonic fractions or concentration parameters.  We prefer
to consider the large spin parameter hypothesis because discs with
large spin parameters are expected to exist since the spin parameter
distribution of figure \ref{figlam} does present an important tail at
large values, and that these discs with large spin parameter values
were not examined in the grid of models previously published 
(BP2000).

\section{Model results and comparison to observations}

 \label{seccomp}

\subsection{Rotation curves}

\label{rotcurv}

In order to compute the Star Formation Rate (equation \ref{equasfr}),
we need to adopt a rotation curve $V(R)$ in our models. As in BP2000,
it is computed as the sum of the contributions of the baryonic matter
(exponential disc) and of the dark matter halo (non-singular
isothermal sphere). 
The use of an exponential disc is justified by the fact 
that 80 \% of the galaxies of the O'Neil sample
(the larger) are well fitted by an exponential disc
(see also Bell et al., 2000).
The shape of the halo used to compute the rotation curve is that of a
 non-singular isothermal sphere. We take the core radius to be a
 constant fraction of the scalelength of the disk obtained by equation
 \ref{equascal1}. The CDM halo density profiles are known to be
 too "cuspy" with respect to observations (e.g. de Blok et al. 2001).
 It is thus difficult to decide which shape to use in the models. 
 Our aim is not to include the most realistic profile of dark matter haloes, 
 but to end up with rotation curves in rough agreement with the observed
 ones (this is actually all that matters for our simple models
 through the equation \ref{equasfr} used to compute the SFR). We 
 make such a comparison in the following.

The shape and amplitude of the resulting curves depend on
the spin parameter and the circular velocity, as a result of the
scaling relations.

It is often argued that LSBs are dominated by dark matter
(e.g. McGaugh et al. 2001, and references therein), and, indeed, in
our models the contribution of the baryonic disc to the total rotation
curve is lower for LSBs than for HSBs, as a result of the lower
densities in the disc.

\begin{figure}  
\includegraphics[width=0.5\textwidth]{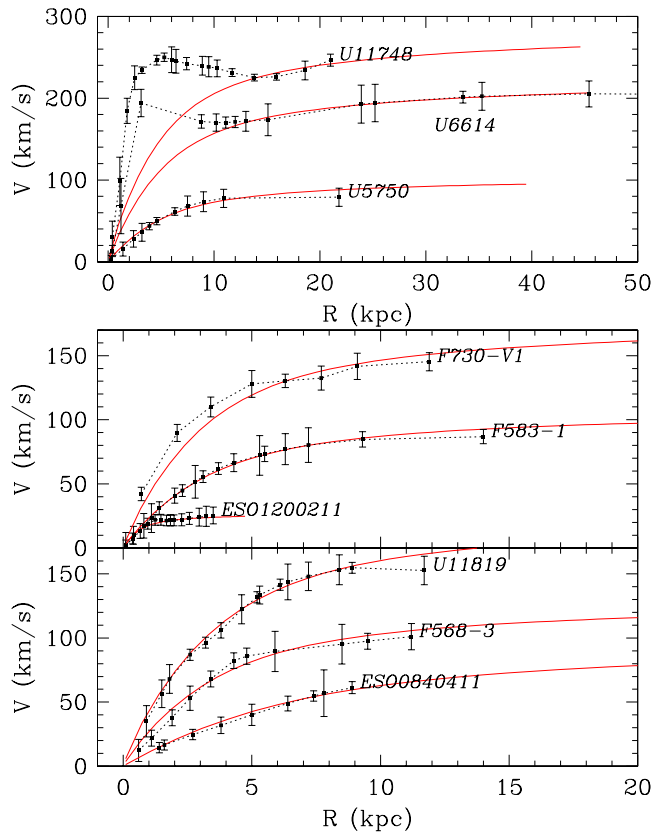}
\caption{
\label{figrotc}
Rotation curves of LSBs, compared with those adopted  in our simple models. 
{\it Dotted curves} and {\it data points}: 
smoothed hybrid H$\alpha$+\HI\ rotation curves derived by De Blok et al. (2001);
{\it Solid curves}: model rotation curves computed with large spin parameters and
appropriate circular velocities. Such rotation curves were used to compute the 
Star Formation Rate (equation \ref{equasfr}) in our models.} 
\end{figure}

To check that our rotation curve are realistic enough for our
purpose, we compare in figure \ref{figrotc}
rotation curves computed with spin parameters larger than 0.07 to a
number of typical rotation curves constructed by de Blok, McGaugh \&
Rubin (2001) for 26 LSB galaxies with available high quality H$\alpha$
and \HI\ data.

For most galaxies a reasonable agreement is obtained between model
curves and observations.  We note that in 2 out of the 26 objects
(U6614 and U11748), a peak appears close to the centre, which may be
due to a pronounced bulge component. Since our models concern discs
only, we cannot reproduce such features and we clearly underestimate
the rotation velocity (and hence the SFR) in the inner $\sim$10 kpc of
these two objects.  Still, these are large and massive galaxies and
only their inner parts are affected.

We consider this comparison as satisfactory and 
we conclude that our simple rotation curves are
realistic enough to compute the SFR with equation \ref{equasfr}.

\subsection {Model results}

\begin{figure}  
\includegraphics[width=0.5\textwidth]{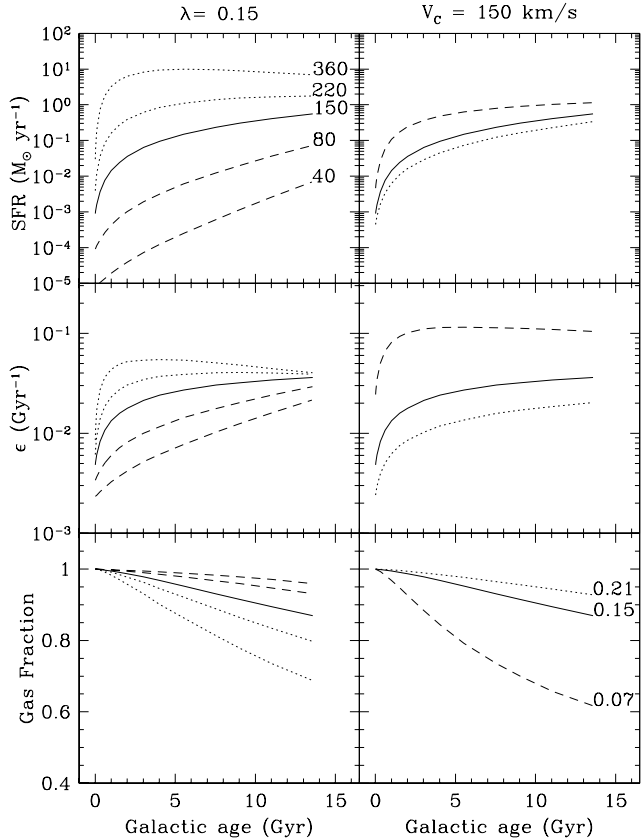}
\caption{\label{figtn}
Some properties of our models as a function of time. From top to bottom: 
Star formation rate, star formation efficiency, and gas fraction.
{\it Left: } results for a mean value of the spin parameter $\lambda$=0.15 and 
for five values of the rotational velocity $V_C$ of our grid of models, as indicated
in the top left panel. {\it Right}:results are given for a typical value of the rotational 
velocity $V_C$=150 km/s and for three values of the spin parameter $\lambda$
as indicated in the bottom right panel.
} 
\end{figure}

\begin{figure}  
\includegraphics[width=0.5\textwidth]{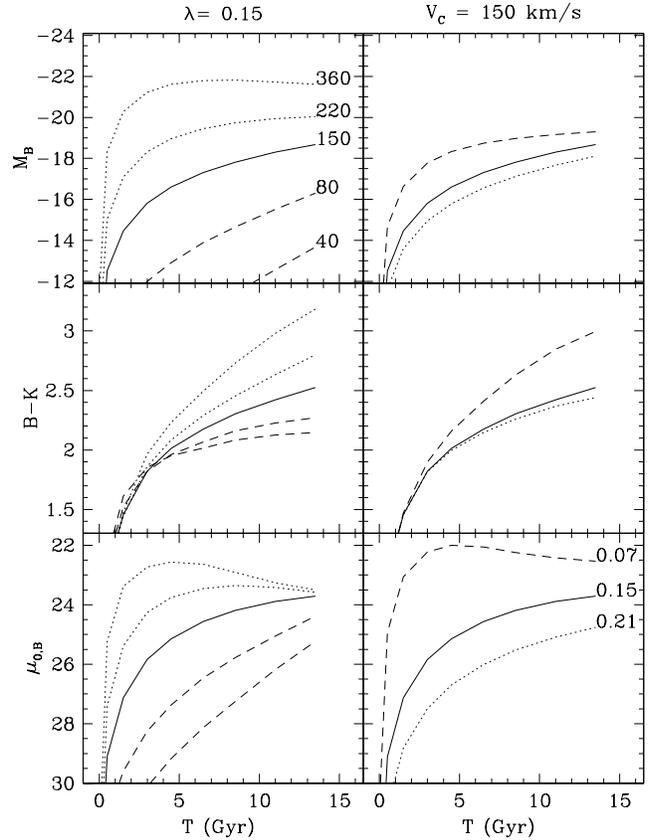}
\caption{\label{figphottime}
Some properties of our models as a function of time. From top to
bottom: B magnitude, B-K colour, and central surface brightness.  {\it
Left: } results for a mean value of the spin parameter $\lambda$=0.15
and for five values of the rotational velocity $V_C$ of our grid of
models, as indicated in the top left panel. {\it Right}:results are
given for a typical value of the rotational velocity $V_C$=150 km/s
and for three values of the spin parameter $\lambda$ as indicated in
the bottom right panel.}
\end{figure}

\begin{figure}  
\includegraphics[width=0.5\textwidth]{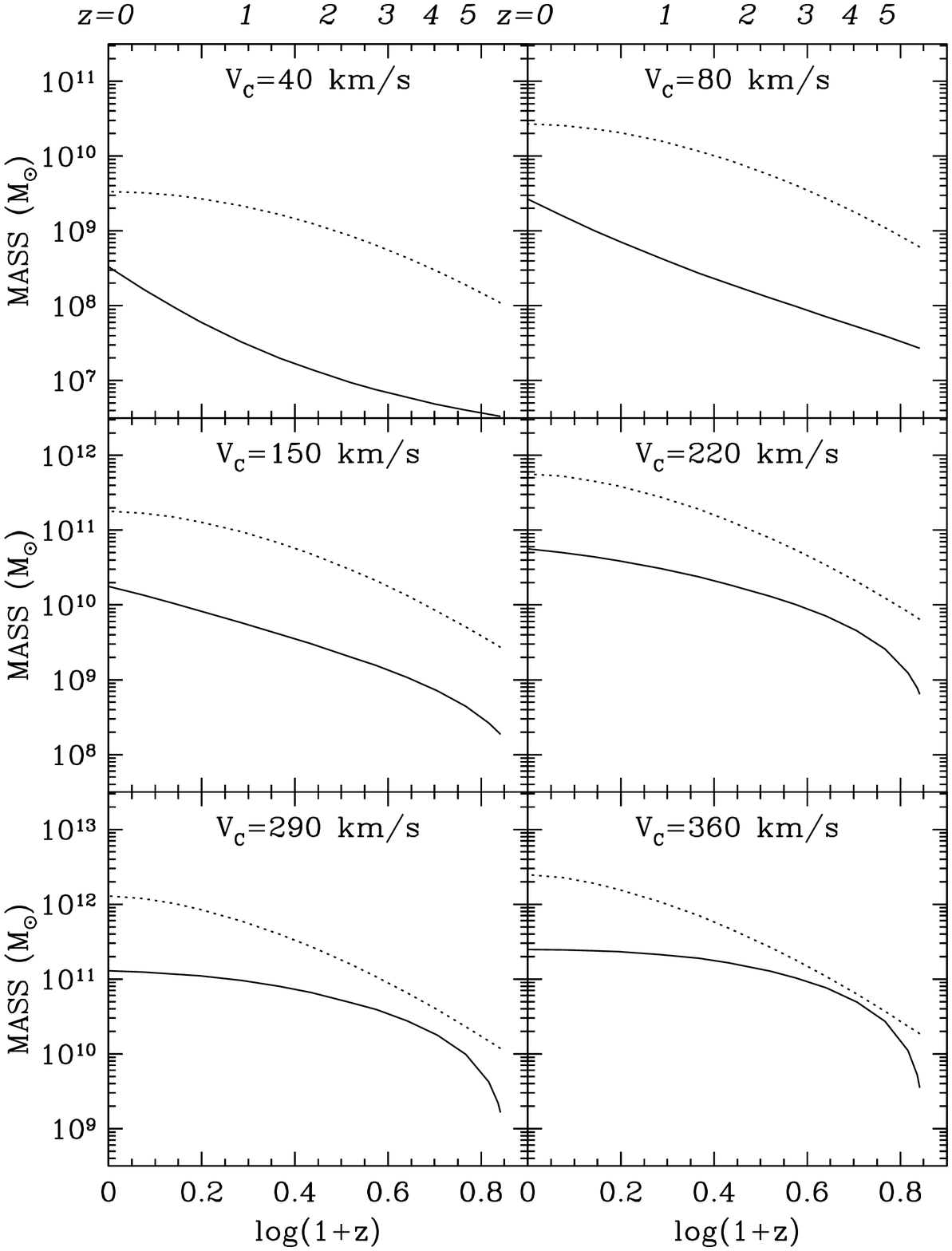}
\caption{\label{figmhist}
Total mass accreted on galaxies of various velocities as 
a function of redshift. The solid line indicates the mass
of baryons in the disk of our models. The dotted line indicates
the history of the dark matter halo ten times as massive as
the disk at the end of the history, according to the universal
mass accretion history of van den Bosch (2002). A redshift
scale is indicated at the top.}
\end{figure}

Figure \ref{figtn} to \ref{figmhist}
display some of our results concerning respectively the
chemical and the spectro-photometric evolution of LSBs. On the left side
of each figure, results are given for a typical value of the spin
parameter $\lambda$=0.15 and for five values of the rotational
velocity $V_C$ of our grid of models. On the right side results are
given for a typical value of the rotational velocity $V_C$=150 km/s
and for three values of the spin parameter $\lambda$ of our grid of
LSB models.

In Figure \ref{figtn}, we present from top to bottom the SFR, the Star Formation Efficiency
(SFE) and the gas fraction. The trends obtained in the case of HSBs (BP2000)
are also reproduced here: more massive discs formed their stars
earlier, had higher star formation efficiencies in the past and have
smaller gas fractions than their lower mass counterparts; these
features characterise also discs with low angular momentum.
LSBs are however different from HSBs: 
the models have lower SFE because of the lower gas densities. As a result,
LSBs at the current epoch are less evolved, with large gas fractions and
young stellar populations.

In Figure \ref{figphottime}, the magnitude (top), B-K colour (middle) and central
surface brightness (bottom) are presented.  An important property is
again similar to what was found in HSBs: the more massive galaxies
(larger rotational velocities) are redder than low mass ones.  This
figure shows also that LSB galaxies were generally fainter in the
past, and that their surface brightness has stayed low during
their evolution, according to the models.

The mass accretion history of our models depends on the mass
of the galaxy. This dependence was constrained by observations
in spirals (e.g. Boissier et al., 2001): it is purely empirical
and is not derived from a cosmological context.
In figure  \ref{figmhist}, we  compare the obtained histories with
the predictions for dark matter halos in our hierarchical universe
with the cosmology $\Omega_M$=0.3, $\Omega_{\Lambda}$=0.7, h=0.65). 
We assume for
each disc that the cold dark matter halo is ten times as massive as
the baryonic disc (at redshift 0), and that
the halo follows the ``universal mass accretion history''
of van den Bosch (2002), depending mainly on the mass of the halo.

For intermediate circular velocities, the mass history empirically
established follows a similar trend as the one of the dark matter
halo. For lower masses, we clearly observe that the baryon accretion
in our models occurs later than the dark matter accretion. This 
might be due to feedback issues in low-mass galaxies
(e.g. Dekel and Woo, 2002). In the simple
models, no feedback is included explicitly, and the mass accretion
rates we were conducted to adopt mimic its effect.

When we move towards more massive galaxies, our baryonic discs
build up more and more rapidly, and for the most massive galaxies
even more rapidly than the dark matter halo. 
It has been already noticed that the properties of massive
spirals (red colours, high metallicities, low gas fractions)
are well reproduced in the simple models by assuming very high
redshift formation. Some more realistic cosmological models might
be useful to study this paradoxal situation, but it is beyond 
the scope of our paper. Note that very massive galaxies represent
a small fraction of the galaxies we are studying in this paper.

\subsection{Tully-Fisher relation}

\begin{figure}  
\includegraphics[angle=-90,width=0.5\textwidth]{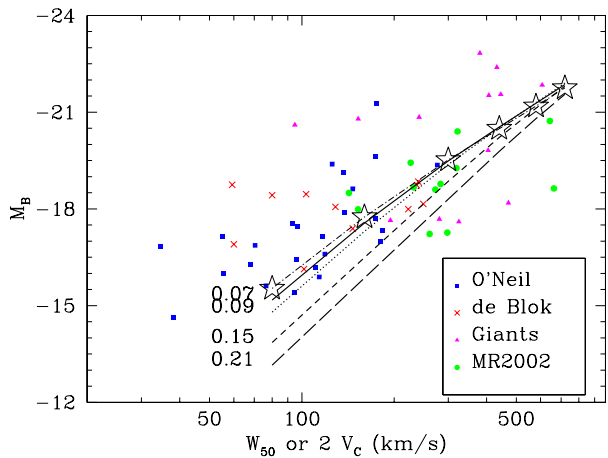}
\caption{\label{figtf}
The Tully-Fisher relation for different samples of LSBs, as indicated
in the plot (see section 2 for details on the data).  The curves
correspond to models with various spin parameters indicated in labels.
A fifth curve marked by stars shows a model with $\lambda$=0.05, i.e. 
corresponding to HSBs.
}
\end{figure}

The Tully-Fisher relation, the linear relation between the absolute
magnitude and the logarithm of the inclination-corrected \HI\ line
width, is one of the best-known properties of spiral
galaxies. Observationally it is still difficult to establish if LSBs
follow the same Tully-Fisher relation as HSB spirals or not.  Some
studies conclude that LSBs follow the same relation as classical spirals
(Zwaan et al. 1995; Sprayberry et al. 1995; Chung et al. 2002), while
in others LSB are found to be underluminous for a given line width
(Persic \& Salucci, 1991; Matthews et al. 1998). Chung et al. (2002)
suggest that some of these differences may be due to selection
effects.

Although O'Neil et al. (2000a) find that a few of their LSB galaxies show large departures
from the standard Tully-Fisher relation, subsequent VLA \HI\ imaging of four of them  
and a new analysis of the single-dish data (Chung et al. 2002) show that all (or most) 
of their original \HI\ profiles may have been contaminated by \HI\ from a nearby 
galaxy. 

In figure \ref{figtf} we plot the absolute blue magnitude as a function of the
\HI\ line-width, corrected for inclination, for the LSBs samples of De
Blok et al. (crosses), O'Neil et al. (triangles), Giant LSBs
(triangles) and the infrared-selected galaxies (filled dots).  The
data for the O'Neil et al. galaxies were taken from Chung et
al. (2002), after cleaning from known contaminations. The combination
of the different sample shows the existence of a Tully-Fisher
relation, though with a considerable scatter. Our LSB models are
overplotted, with each curve corresponding to a different spin
parameter. Models with a lower spin parameter are also shown (stars)
for reference. 
It is assumed that the circular velocity of our models,
$V_C$, corresponds to half the inclination-corrected \HI\ line-width,
0.5W$_{50}$/sin(i).

The models predict the existence of a Tully-Fisher relation
for both HSB and LSB disc galaxies. However, the adopted scaling relations
(section 3.1) lead to a difference between the HSB and LSB TF
relations: at a given rotational velocity the luminosity of LSBs is
lower than for HSB spirals because of differences in their star
formation history and gas fraction. The amplitude of the effect
depending on the velocity $V_C$, the slope of the relation changes
with the spin parameter.

Although the modeled and observed LSB Tully-Fisher relations show
relatively similar slopes, the model values lie on average about 2 mag
below the mean observed absolute magnitudes. This shift could be due
to systematic differences between the measured and modeled circular
velocities and to uncertainties in both of them.

\subsection{Scalelength and central surface brightness}

\begin{figure}  
\includegraphics[width=0.5\textwidth]{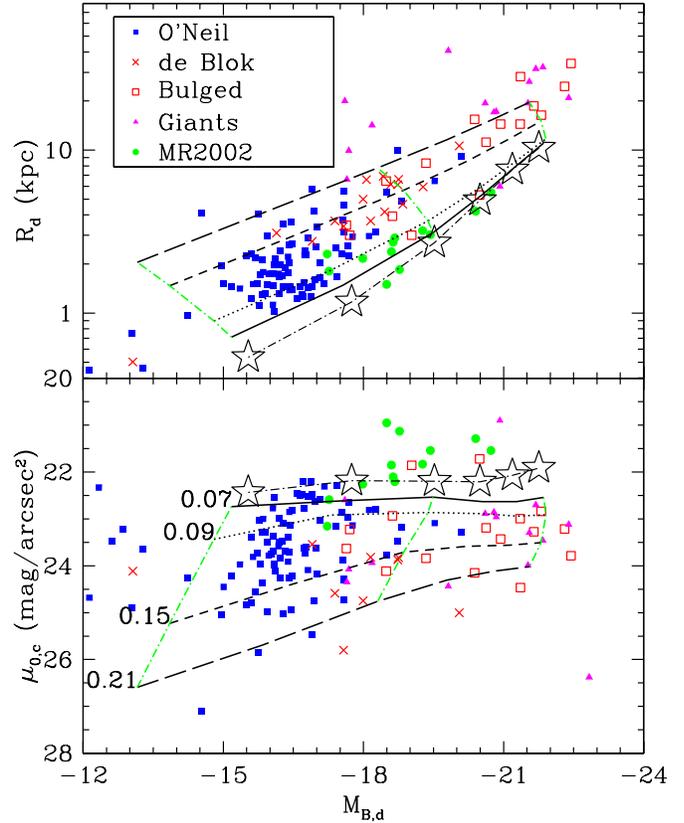}
\caption{\label{figmur}
$B$-band disc scalelength (top) and central surface brightness
(bottom) deduced from fits to the surface brightness profiles as a
function of the absolute blue magnitude of the disc component.
Plotted are LSBs from the samples of O'Neil et al. (filled squares),
de Blok et al. (crosses), Giant LSBs (triangles), Beijersbergen (open
squares) and the infrared-selected galaxies (filled dots).  Curves
correspond to models with different spin parameters (as indicated in
the bottom panel).  The curve marked with stars correspond to HSBs
($\lambda=0.05$), while the others correspond to LSBs.  The grey lines
join models with the same circular velocity (from left to right: 40,
150, and 360 km/s, respectively.  }
\end{figure}

Figure \ref{figmur} displays  the disc  scalelength and central surface
brightness in the B-band as a function of $M_{B,d}$, the absolute magnitude of the disc component
(as defined in section \ref{secmagalpha}).
>From the adopted scaling relations (section \ref{secscal}), we expect 
that the scalelength will increase with luminosity, while the central surface brightness will
depend more strongly on the spin parameter. This is indeed the case, as seen
in figure \ref{figmur}.

We note that the lack of observed galaxies in some parts of these diagrams 
(specifically, those with large luminosities but very faint central surface brightnesses,
the so-called ``Malin 1 cousins'') might result from selection effects. 
On the other hand, in view of  
the distribution of the spin parameter (figure \ref{figlam}) and of
the circular velocity function (Gonzalez et al. 2000) it is expected that 
only a small number of LSB galaxies with large rotational velocities should exist. 
Therefore, one would expect such galaxies to be quite rare compared to LSBs 
of considerably smaller size or less extreme surface brightness.  

Our models  produce LSB disc galaxies similar in size and surface brightness 
to those of the presently available observational samples. However,  within
the framework of our models, giant discs  with lower surface brightness could 
in principle be produced by systems with even larger $\lambda$ values than those 
adopted in our present calculations.

Finally, the scaling relationships indicate that LSBs should have larger 
scalelengths than HSBs for the same mass. The models for HSBs (stars) in
Figure \ref{figmur} are indeed located at smaller scalelenths than the models for LSBs.
The same result is obtained in the observations since the MR2002 sample is
characterised by smaller scalelengths since the other data, and correspondingly
higher surfaces brightnesses.


\subsection{Gas content}

\begin{figure}  
\includegraphics[width=0.5\textwidth]{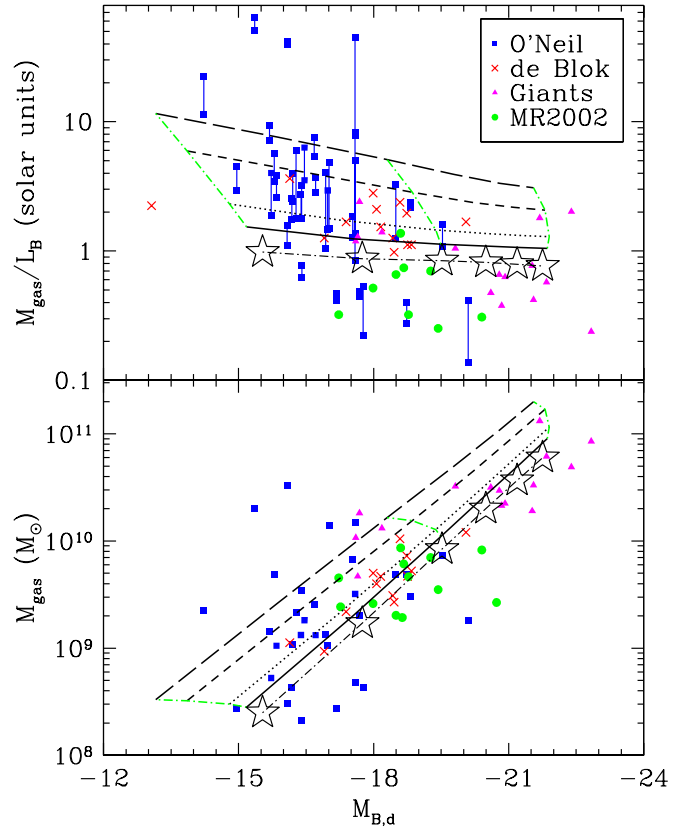}
\caption{\label{figgas}
Gas mass-to-blue light ratio (top) and gaseous mass (bottom) as a function
of the magnitude of the disc component. 
Dark curves correspond to models with different spin parameters (same as in previous figures).
The grey lines join models with the same circular velocity (40, 150, and 360 km/s).
}
\end{figure}

Our models predict a correlation between the absolute disc magnitude
and the gaseous mass (bottom panel of figure \ref{figgas}).  This
results mainly from the mass-luminosity relation: more massive
galaxies are more luminous \emph{and} contain more gas (as expected if
the gas-to-total mass ratio does not depend strongly on
luminosity). The data follow the same trend, but the observational
scatter is larger than what our models can account for, especially in
the faint LSB range (where most of the data come from the O'Neil et
al. sample).

Note that in a recent study, Galaz et al. (2002) presented a clear
correlation between the HI mass and the Ks magnitude for a sample of
galaxies they studied with near-infrared imagery, confirming the trend
we observe in Figure \ref{figgas}.  As a direct consequence of the scaling
relationships, using the words of Galaz et al. (2002), ``bigger
galaxies have more of everything''.

The gas mass-to-light ratio ($M_{gas}/L_B$) is an interesting
intensive quantity, more related to the ``chemical
state'' of the galaxy than the absolute gas mass. Our models predict
larger values of $M_{gas}/L_B$ for LSBs than for HSB spirals, as a
result of the lower star formation efficiency in the former.  In
figure \ref{figgas} it appears that our models predict a slightly
declining trend with absolute magnitude: more massive galaxies have
proportionally smaller gas fractions, due to the smaller time-scale of
infall rates.  The data show a steeper trend (arising mainly from the
giant LSBs) with a broader dispersion than allowed by our models.  For
the LSBs data of O'Neil et al., two values of this ratio are shown,
giving an indication of the corresponding uncertainty: one value is
computed with a luminosity derived from the absolute magnitude
$M_{B,d}$ of the disc only and one is based on the total absolute
magnitude of the galaxy (the two {\it filled square} linked by a
vertical line).

Note that the HSBs models have lower gas contents than the LSBs models,
what is in agreement with the observations: the MR2002 sample is
characterized by smaller values of the gas to luminosity ratio than
the other samples.

A way to obtain a stronger decrease of $M_{gas}/L_B$ with $M_{B,d}$ in
our models would be to introduce a different parameterisation of the
key physical ingredients in order to consume more gas in massive
discs, either by either making the gas available earlier (more rapid
infall) or by changing the prescription for star formation (enhancing
the SFR efficiency).  Still, given the large observational
uncertainties and the crudeness of our ``1st order'' models, we
consider that the simple ``large-spin'' models presented here
reproduce satisfactorily the observed $M_{gas}/L_B$ values.  We also
note that departures from these values could be due to star formation
events that occur in addition to the smooth history of the simple
models, as will be discussed in section \ref{burstanti}.

\subsection{The luminosity-metallicity relation}

\begin{figure}  
\includegraphics[angle=-90,width=0.5\textwidth]{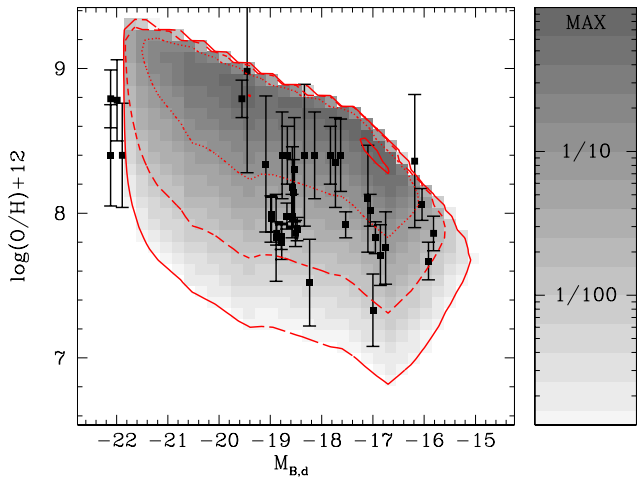}
\caption{Luminosity-metallicity relation: the data correspond to the abundances of oxygen in 
H\,{\sc ii} regions as a function of the absolute magnitude of the
parent LSB galaxy (McGaugh 1994). The contours and greyscale levels
show the expected model abundances, weighted by the velocity
distribution, the spin distribution, and the star formation rate (see
text).
\label{figox1}}
\end{figure}

The ``luminosity-metallicity'' relation is an important property of
(HSB) spiral galaxies (e.g. Zaritsky, Kennicutt \& Huchra 1994):
massive galaxies contain proportionally more metals than low-mass
ones. The metallicity is estimated from the oxygen abundance at a
given radius in the disc of the galaxy (see Zaritsky et al. 1994 for
details).
 
This relation is very important for studies of galactic evolution
since it suggests that massive galaxies have processed more material
into stars, or that low mass galaxies have lost part of their metals
or a combination of both occurred.  The former hypothesis was adopted
in Boissier et al. (2001), on the basis of observed star formation
efficiencies and gas fractions, for spirals with rotational velocities
as low as 80 km/s.  Recently, Garnett (2002) used a large data sample
of spiral and irregular galaxies with rotational velocities as low as
5 km/s and convincingly argued that below $\sim$120 km/s the effective
yield of galaxies is reduced, most probably as a result of mass
loss. We note that this effect, by itself, is not sufficient to
explain the observed colour-luminosity relation of spirals, with the
most massive ones being redder than their low mass counterparts; a
younger age for the latter has to be assumed in order to account for
the data (see Prantzos 2002 for a recent review).

Because our LSBs models are an extrapolation of the HSB ones, we expect them to produce a 
luminosity-metallicity relation in the case of LSBs also. However,
McGaugh (1994)  compiled metallicity measurements in LSBs and concluded that 
they do not display such a relation [his data are shown in figure \ref{figox1}
as a function of $M_{B,d}$]. It should be noted that
McGaugh (1994) excluded giant LSBs from his analysis, while we consider a continuous range 
of galaxies, from dwarfs to giants; if we add the giants to the sample, a small trend is observed
(see figure \ref{figox1}). On the other hand,
the abundance measurement in LSBs concern only a few H\,{\sc ii} regions per object, 
located at various galactocentric radii. Therefore, it is impossible
to determine an abundance at a ``characteristic'' radius and the situation is very different
from the work of Zaritsky et al. (1994) concerning HSB spirals. This difficulty 
introduces a large scatter in the luminosity-metallicity relation (if there  exists one), 
compared  to the case of  the HSBs. 

In figure \ref{figox1} we compare the data of McGaugh (1994) with the metallicities of our 
LSB models. Since the observations were done at various radii, for a comparison with 
the models we computed the distribution expected in the absolute magnitude - log(O/H) plane, 
by weighting the results with the distributions of the spin parameter 
(figure \ref{figlam}), 
circular velocity (we adopt the distribution of Gonzalez et al. 2000), 
and Star Formation Rate. The reason for introducing the latter is that
abundances are measured in  H\,{\sc ii} regions, implying the existence
of massive stars and, hence, important star formation activity.

As can be seen from figure \ref{figox1}, the abundances measured in
LSBs are in satisfactory agreement with the values expected from our
simple models, at least in a statistical sense.

Galaz et al. (2002) argue that the J-Ks colour index they measured in
their near-infrared galaxy sample is a good indicator of the
metallicity. On this base, they deduce the existence of a
luminosity-metallicity relationship, the redder colours corresponding
to the galaxies with the largest stellar mass. From the variation of
the J-Ks index between the less luminous and the more luminous
galaxies, they conclude that the metallicity increases by a factor
between 20 and 100 between them. In a logarithmic scale,
this is 1.3 to 2 dex, what compares well with the slope we can estimate
from Figure \ref{figox1}.

\label{massmet}

\subsection{Colours}

\begin{figure}  
\includegraphics[width=0.5\textwidth]{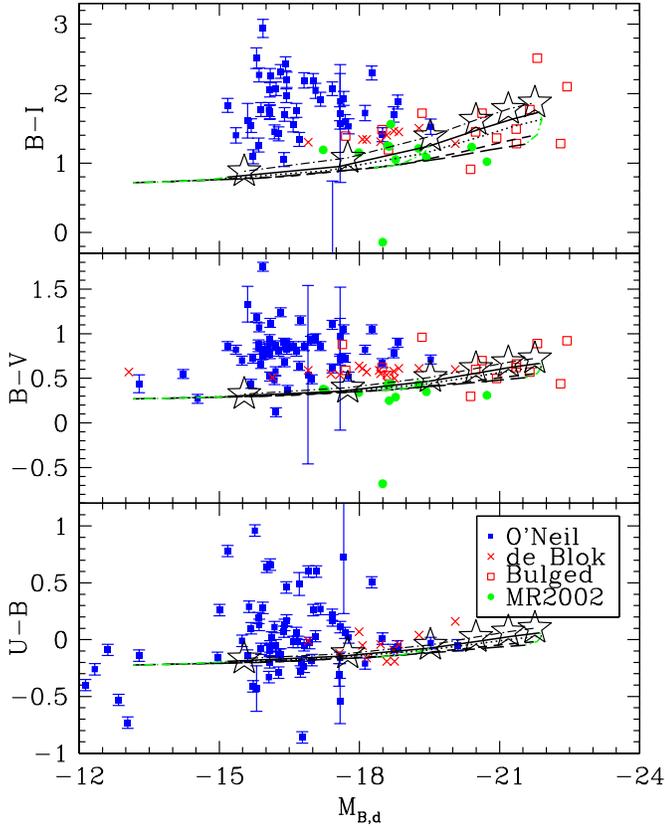}
\caption{Colour indexes as a function of blue magnitude.
\label{figcol}
Curves correspond to models with various spin parameters like in
previous figures.
}
\end{figure}

Three colour indices are presented as a function of absolute disc
magnitude in figure \ref{figcol}. The models are characterized by blue
colours and an extremely small colour dispersion, especially for
low-mass discs. Both features stem from the fact that most of the
model galaxies have formed the bulk of their stars toward the end of
their history (see figure \ref{figtn}).  The differences in their early star
formation histories play a minor role (especially in low-mass discs)
because the colours are always dominated by the youngest population of
stars.

Although these model results are in broad agreement with the ``blue''
LSBs of the observational sample, they do not reproduce the colours
observed in the large number of ``red'' LSBs.  It is worth noting that
the red LSBs are mainly found in the O'Neil et al. (2000) sample, part
of which consists of cluster galaxies, though it should also be noted
that O'Neil et al. (2000a) find no difference between the properties
of their LSBs in clusters and in the field.

We also point out that colour uncertainties may be larger
than usually quoted ($\sim 0.1$ mag), since O'Neil et al. (2000a) and
Bell et al. (2000) find for the same object (P1-7) $B-V \sim 0.9$ and $B-V=0.6 \pm 0.1$, 
respectively.  Notwitstanding this difference and the importance of the
uncertainties in LSBs' colours, the existence of ``red'' LSB discs is undisputable and
that property is clearly  not
reproduced by the simple models presented here.  In section
\ref{burstanti}, we suggest that another ``ingredient'' is necessary to reconciliate 
the simple models with the existence of such red galaxies.

\subsection{Summary of the ``simple'' model calculations}

The results of our simulations  can be summarised as follows:

$\bullet$ Simple models (i.e. with smooth SF histories), differing by HSBs 
only by their  larger  spin parameters ($\lambda$), 
are in reasonable agreement with several observed
properties of LSBs, e.g.  the relations between absolute disc magnitude and central 
surface brightness, disc scalelength, metallicity, as well as total gas content. 
The models also predict a Tully-Fisher relation, in broad agreement with observations.

$\bullet$ Some important discrepancies exist between the observations and 
the ``simple'' models: the scatter in many of the observed relations (Tully-Fisher, 
gas mass vs. luminosity, colours) is much larger than predicted by the models. 
In particular, the ``red'' LSB galaxies are not reproduced by any of our models, since
``by construction'' they are dominated by the young stellar population.
 
The discrepancies between models and observations are particularly
important.  In fact, the models of BP2000 (on which this work relies)
were developed mainly for field galaxies with smooth star formation
histories. If the environment of LSB is different or if they are more
sensitive to perturbations (because of their lower densities), some of
their properties may be due to departures from the ideal smooth star
formation histories. This idea is tested in the next section, where we
try to reconcile the large-spin models and the discordant observations
by adding ``events'' (bursts and truncations) on otherwise smooth star
formation histories. Indeed, O'Neil et al. (2000a) suggested that the
properties of both blue and red LSB galaxies can be reproduced by
population synthesis models, assuming ``on'' and ``off'' states of the
star formation for different stellar populations.  Van den Hoek et
al. (2000) already suggested that small amplitude bursts may play a
role in determining the colours of LSBs. In their N-body simulations,
Gerritsen \& de Blok (1999) argued that Star Formation Rate
fluctuations are responsible for the large range of colours observed in
LSBs and predicted that less than 20\% of gas-rich LSBs should be
``red''.

\section{Starbursts and truncated star formation histories}

\label{burstanti}

\begin{figure}  
\includegraphics[width=0.5\textwidth]{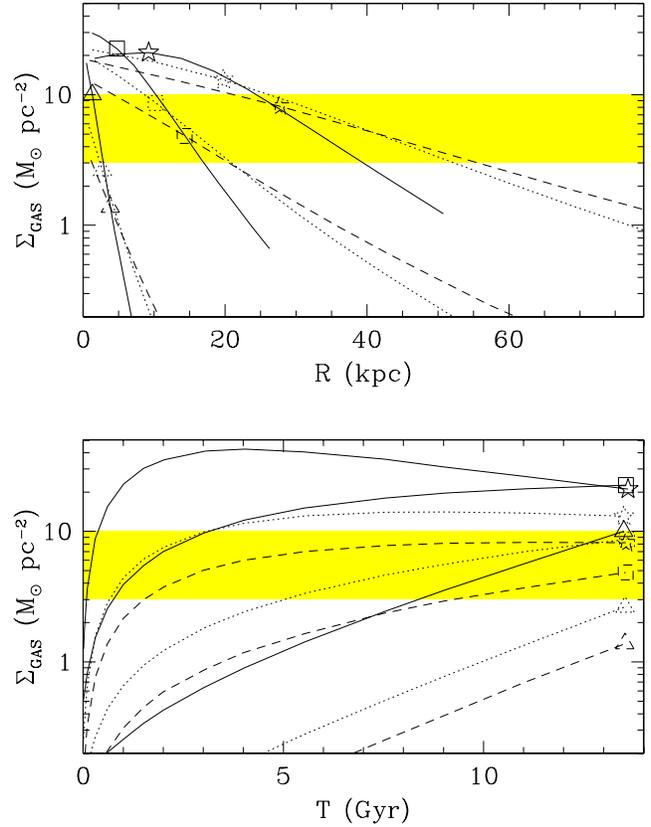}
\caption{The gas surface density in a few models as a function of radius (top) and its 
evolution with time at a radius of one scalelength (bottom). 
The solid/dotted/dashed curves correspond respectively to $\lambda$=0.07/0.15/0.21. 
The shape of the symbol on each curve indicates the velocity: 40 km/s (triangles),
150 km/s (squares) and 290 km/s (stars). In the top panel, the symbol is placed
at a radius equal to one scalelength. 
The grey  area indicates the ``threshold'' of star formation in disk, according to
observations by Kennicutt (1989, see also section 5)
\label{figthres}
}
\end{figure}

In this section, we investigate a way to produce red LSB galaxies
while introducing only small changes to the ``simple'' models
presented in section \ref{models}.  We will refer to the star
formation histories derived with the simple models as ``standard''
histories, which represent a smoothed average of the true evolution of
the LSBs, to which we will now add star formation ``events''.

First, we study the effect of a starburst added to the ``standard'' SF
history of one of our models, and then we consider the effect of a
truncation of the ``standard'' star formation at a recent epoch.

There are reasons to believe that the star formation rate behaves in 
such a way in LSB galaxies.

Martin and Kennicutt (2001) show that in nearby spirals, beyond
the ``threshold radius'', where the density of gas becomes smaller
than the critical density (according to the Toomre instability
criteria) a few HII regions (and then recently formed stars) are 
still found.  They argue that local perturbations and
density waves can temporarily and locally enhance the density to
values high enough to allow star formation. This way of forming
stars is obviously less smooth than the one described in the
simple averaged models.
Despite a large scatter, we can derive from the figure 9 of Kennicutt (1989)
that the critical gas density at which the threshold is found 
range between 3 and 10 $M_{\odot} pc^{-2}$ for most of the galaxies.
This range of values is indicated in the Figure \ref{figthres} as a grey area.
The gas surface density in some of our LSB models is indicated as 
a function of radius at the present epoch (top panel). In the bottom
panel, each curve presents the evolution with time of the gas surface
density at a radius equal to one scalelength, for the same models as
in the top panel.

Obviously, the gas surface
density has been lower than (or close to) the threshold value observed in
spiral galaxies, in most of our model galaxies and most of the time.  
Thus, the star formation is more likely to
proceed through
a succession of quiescent phases (when the density is lower than the
threshold), and relatively active phases (when enough gas has been
accumulated by infall and condensed by local perturbations or spiral
modes to reach the threshold). This scenario is indeed
suggested by Galaz et al. (2002) in order to explain the scatter in optical
data by a sporadic past star formation history.

Another possibility is that the star formation history is affected by
feedback from supernovae.  Feedback effects may be crucial to 
the history of LSB galaxies (see for instance the study of
Dekel and Woo, 2002).  We can speculate that the effect of supernovae
could be similar to the empirical threshold discussed above: stopping
the star formation by blowing the gas away and delaying the birth of the
next generation of stars until enough gas has settled again in the disk.

It is beyond the capacities of the simple models to test the details
of these ideas concerning the way the star formation rate is affected
in low density environment
by threshold effects, spiral modes, perturbation, feedback. 
This task could be undertaken only with much
more sophisticated hydrodynamical models (e.g. Samland and Gerhard, 2003).
We compute instead 
two simple examples for illustrative purpose.

Of course, the star formation history of a galaxy may consist of a
series of ``bursts'' and ``drops'' in its star formation rate. 
However, the
``chemical'' properties of a 13 Gyr old galaxy with such a jittery SFR
would be virtually identical to those resulting from an averaged star
formation history, while its colours may be affected mainly by the
last of such events. Therefore, our adopted ``smooth plus last event''
model, while simplistic, may be a good enough approximation for the
present study of LSBs.

\subsection{Adding a star burst}

\begin{figure}  
\includegraphics[width=0.5\textwidth]{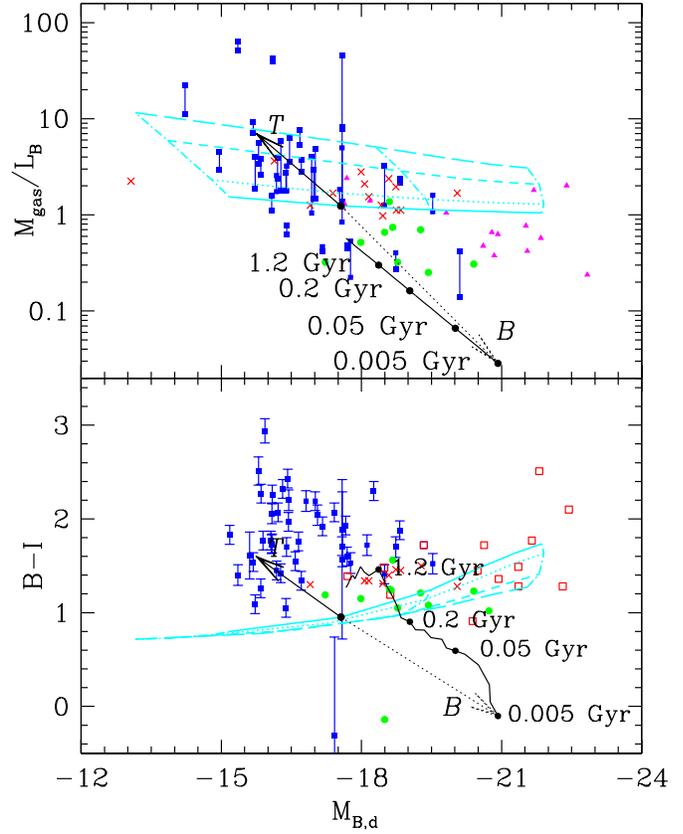}
\caption{
Effects of star bursts and truncations of the star formation on the
gas mass-to-light ratio in the $B$-band (top panel) and $B-I$ colour
(bottom panel). The data and the model grids are identical to those in
figure \ref{figcol} and \ref{figgas} (except that 	we do not reproduce
the HSB models).  The effect of a star burst in
addition to a ``simple'' model ($\lambda$=0.07, $V_C$=80 km/s) is
shown by the arrow labelled ``B'' (at an age of 0.005 Gyr).  The curve
towards the tip of the arrow corresponds to increasing burst ages
along the curve. The effect of the truncation on the star formation
history of the same model one Gyr before the present epoch, is shown
with the arrow labelled with a ``T''.  
\label{figbtt}}
\end{figure}

Figure \ref{figbtt} shows the $B-I$ colour and the gas mass-to-light ratio in the $B$-band
for the galaxies from the various samples and the grid of ``standard'' models.

It would be difficult to get a clear insight by adding bursts of
various ages, intensities, metallicities, etc.  to all of the
``simple'' models. Instead, we prefer to study one illustrative
example, by adding a single burst of star formation to the
``standard'' star formation history of one of our models
($\lambda$=0.07, $V_C$=80 km/s). This burst creates 5 10$^{6}$
M$_{\odot}$ of stars with a metallicity equal to one third of solar
and a ``normal'' IMF (from Kroupa et al. 1993 in all our models).

When the burst is 0.005 Gyr old, the model results correspond to the
position indicated by the tip of the arrow labelled ``B'' in 
figure \ref{figbtt};
for older bursts the results are located along the curve, as indicated
by the corresponding numbers, and they are progressively closer to
those of the ``standard'' (smooth SF) models.
 
For very recent bursts, it seems likely that the brightest part of the
galaxy would be the burst itself, in which case the galaxy would
probably not be classified as an LSB, but rather as a ``blue compact''
(if the burst is centrally located). Legrand et al. (2000) proposed
such a scenario for {\sc i} Zw 18, a blue compact galaxy of very low
metallicity, which could result from a star burst occurring in an LSB
galaxy after a long period of low and constant star formation rate.

For burst ages under 0.5 Gyr, the colours of the galaxy are bluer than
without the burst; beyond that age the old stars of the burst dominate
the stellar populations of the galaxy, which becomes redder than in
the ``standard'' model.  However, even for a very old burst, we do not
obtain a very red galaxy.  In fact, after 12.5 Gyrs the colours of the
galaxy are quite close to those of the smooth ``standard'' model.

Assuming that all the gas taking part in the burst is removed from the
galaxy, we can also compute a total gas mass-to-light ratio. For young
bursts, $M_{gas}/L_B$ is lower than without the burst (because the
luminosity increases significantly and the gas amount is reduced). For
older bursts, $M_{gas}/L_B$ proceeds towards ``standard'' values, as
the luminosity of the burst (which dominates the stellar population)
decreases with age.

The results obtained here suggest that adding star bursts to the smooth SF history of
the ``standard'' model  can increase the scatter in the 
gas mass-to-light ratio (see figure \ref{figbtt}); 
it will also  increase the scatter around the 
Tully-Fisher relation, especially for luminous LSBs with a relatively 
low gas mass-to-light ratio.

\subsection{Truncating the star formation}

\begin{figure}  
 \includegraphics[width=0.5\textwidth]{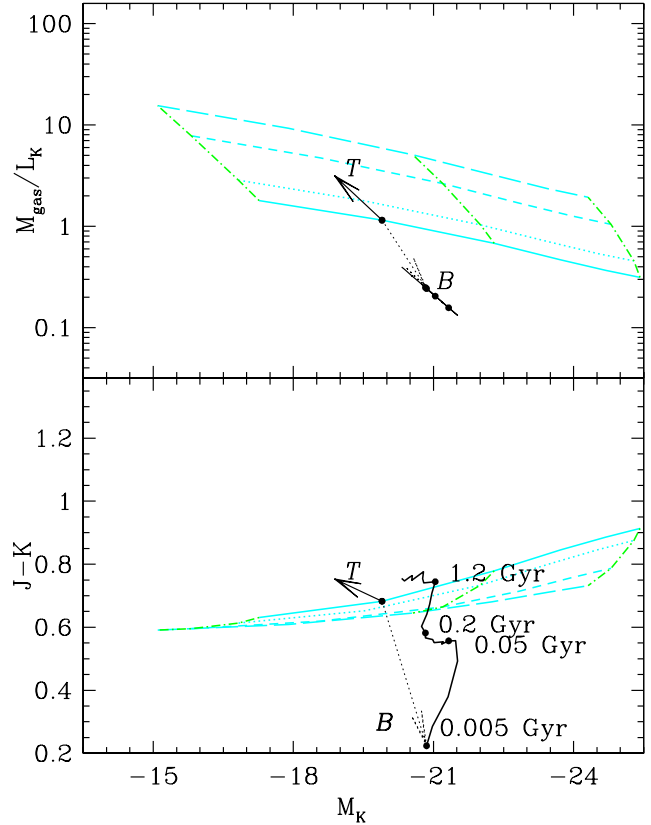}
\caption{Illustration of the effects of star bursts and truncations of the star formation 
on the gas mass-to-light ratio in the $K$-band  (top panel) and the near infra-red 
$J-K$ colour. A comparison with figure \ref{figbtt} shows that deviations are smaller in 
the infra-red than in the optical bands.
\label{figbttir}}
\end{figure}

To illustrate the effect of truncations of the star formation history,
we added to the same model ($\lambda$=0.07, $V_C$=80 km/s) the
condition that 1 Gyr before the present epoch the star formation
stopped abruptly. As a result, the stellar population reddens
passively, since no young and blue stars are created.  The resulting
stellar population is different from that of a ``normal'' galaxy,
without the need to invoke a special IMF.

The effect of truncating the star formation history in this way is
shown in figures \ref{figbtt} and \ref{figbttir} by arrows labelled with a
``T''. The luminosity decreases by about 1.5 magnitudes due to the lack
of newly formed stars, while the colours become quite red. This model
corresponds rather well to the faint red LSB galaxies of the O'Neil et
al. sample.

Because of the decrease in luminosity, galaxies with larger than
average gas mass-to-light ratio are obtained.  Notice, however, that
the scatter in the $M_{gas}/L_B$ vs $M_B$ relation that is obtained by
adding truncations to a smooth SF history is smaller than that due to
the diversity in the values of the spin parameter $\lambda$.  Another
consequence is that truncations of the SF will increase the scatter in
the Tully-Fisher relation, especially for low-luminosity systems (as
is the case with starbursts)


\subsection{The effects of bursts and truncations}

The two simple examples presented in the previous sections
illustrate that a starburst, or a truncation of the star formation, may strongly 
affect the colours, luminosity, and gas mass-to-light ratio of LSB galaxies, with respect to
the ``simple'' models with smooth histories discussed in section \ref{models}.

If the simple models reproduce the average history of LSB galaxies, then the occurrence of star bursts of 
various ages, masses and metallicities, and of truncations of star formation at various galactic ages, may 
explain the  existence of the ``red'' LSBs as well as the large scatter found in several 
observational properties 
of LSBs, like their Tully-Fisher relation, gas mass-to-light ratio, and colours.

If some LSBs have had recent truncations in their star formation
history their colours may be a poor indicator of their true nature;
indeed, a massive LSB disc (having an old stellar population according
to our models) may have the same $B-I$ colour as a less massive one
with a truncated star formation history, evolving passively for e.g. a
Gyr.  This is in agreement with the finding of Bell et al. (2000) that
``red'' LSB galaxies form an heterogeneous group.

Finally, we note that the events we invoked (bursts and truncations)
should affect more strongly the optical than the infra-red properties
of LSBs, as the latter are more sensitive to the old, average, stellar
population. We present in figure \ref{figbttir} the equivalent of
figure \ref{figbtt} for the near infra-red $K$-band, showing the
effects of the burst and the truncation.  A comparison between figures
\ref{figbtt} and \ref{figbttir} shows that infra-red properties are
indeed less affected, and that we obtain less extreme colours in the
infra-red, indicating that adding star formation events will introduce
a relatively much smaller scatter in the colours, Tully-Fisher
relation and gas mass-to-light ratio in the infra-red than in the
optical.

In their recent study of a near-infrared sample of galaxies, Galaz et
al. (2002) foundd that the near-infrared data and colours are less
scattered than the optical ones. This is in full agreement with our
previous deduction. They also found a decrease of the HI mass to light
ratio with the Ks magnitude, together with an increase of the value of
the J-K colour index, both corresponding qualitatively to the results
of figure \ref{figbttir}.

\section{Conclusions}

\label{findesharicots}

We have compiled data from various observational samples of Low
Surface Brightness galaxies (LSBs) in order to cover the whole
presently known range of their properties, from ``dwarfs'' to
``giants'', from ``blue'' to ``red'' galaxies and from those at the
limits of detectability in central disc surface brightness to objects
straddling the adopted boundary between LSBs and HSBs (High Surface
Brightness galaxies).  We have not included in our study very faint
dwarfs, but only gas-rich LSB discs, with rotational velocities larger
than 40 km/s or so.  We compared them to the results of models of
their chemical and photometric evolution, obtained under the
assumption that these LSBs are similar to HSBs except for a larger
angular momentum.  This was done on the base of successful models of
HSBs that were presented in a recent series of papers.

The comparison is relatively satisfactory: observed disc scalelengths,
central surface brightnesses and total gas masses are compatible with
the model predictions. In their recent work, Schombert et al. (2001)
also found that the gas fraction of their LSB dwarfs is compatible
with our main assumption.  However, the star formation history of some
of the LSBs appears to be different from this simple picture, as
indicated by the large scatter found in several observational
relations (Tully-Fisher, total gas mass-to-luminosity) as well as by
the existence of red galaxies, which cannot be not explained within
the framework of the simplest models (i.e. large angular momentum
discs with a smooth star formation history).

In order to account for the discrepancies between the simple models
and the observations we suggest that star formation ``events'', bursts
and truncations, should be included to the simple model.  We show that
these ingredients may easily reconcile the observations with the
models, since they reproduce the observed large scatter in certain
relations; besides, truncations in the SF history may account for the
existence of the red LSBs found by O'Neil et al. (1997b). Such
``events'' may result from the fact that the gas surface density in
most LSBs is probably lower than the threshold for star formation
found by Kennicutt (1989), or from supernovae feedback (Dekel and Woo, 2002). 
Such events are
however unlikely to alter as much the near infrared data as the
optical ones, quite in agreement with the observations of
Galaz et al. (2002)

We conclude that LSBs may be the equivalent of HSBs with larger
angular momentum (spin parameter $\lambda$), but that their observed
properties could be largely affected by recent events in their star
formation history; in that respect, they present similarities to the
case of blue compact galaxies.


\def\apj{ApJ}
\def\aj{AJ}

\label{lastpage}

\end{document}